\documentclass[11pt]{article}
\usepackage[sfdefault]{carlito} 
\usepackage{geometry}
\usepackage{setspace}
\usepackage{titlesec}
\usepackage{hyperref}
\usepackage{graphicx}
\usepackage{amsmath}
\usepackage{bm}
\usepackage{hyperref}
\usepackage{wrapfig}
\usepackage{xcolor}
\usepackage{authblk}
\usepackage{lipsum}
\usepackage[font=small,labelfont=bf]{caption}

\titleformat{\section}{\normalfont\fontsize{11}{13}\bfseries\sffamily}{\thesection}{1em}{}
\titleformat{\subsection}{\normalfont\fontsize{11}{13}\bfseries\sffamily}{\thesubsection}{1em}{}
\titleformat{\title}{\normalfont\fontsize{14}{16}\bfseries\sffamily}{}{0em}{}

\title{Density-Functional Perturbation Theory with Numeric Atom-Centered Orbitals}
\author[1]{Connor L. Box}
\author[1]{Reinhard J. Maurer}
\author[2]{Honghui Shang}
\author[3]{Matthias Scheffler}
\author[4]{Volker Blum}
\author[5]{Christian Carbogno}
\author[6]{Mariana Rossi\thanks{mariana.rossi@mpsd.mpg.de}}
\affil[1]{Department of Chemistry, University of Warwick, UK}
\affil[2]{Key Laboratory of Precision and Intelligent Chemistry, University of Science and Technology of China, Hefei, China}
\affil[3]{NOMAD Lab, Fritz-Haber Institute, Berlin, Germany}
\affil[4]{Department of Mechanical Engineering and Materials Science, Duke University, NC, USA}
\affil[5]{Theory Department, Fritz-Haber Institute, Berlin, Germany}
\affil[6]{MPI for the Structure and Dynamics of Matter, Hamburg, Germany}
\date{}

\begin{document}

\maketitle

\section*{Summary}

\setlength{\intextsep}{1.0pt}%
\setlength{\columnsep}{8pt}%
\begin{wrapfigure}{r}{0.4\textwidth}
  \begin{center}
    \includegraphics[width=0.25\textwidth]{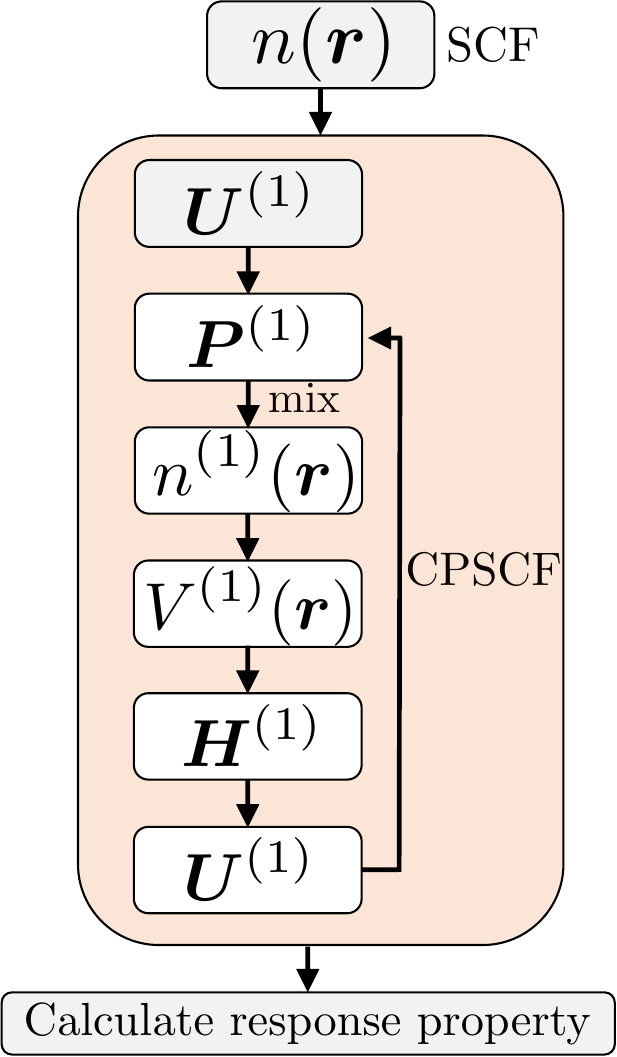}
  \end{center}
  \caption{Self-consistent workflow for response properties in FHI-aims. The ground-state density $n(\bm{r})$ is used for a first guess of the expansion coefficients $\bm{U}^{(1)}$, which enters the response density-matrix $\bm{P}^{(1)}$, which then defines the response density $n^{(1)}$.  $n^{(1)}$ enters the perturbation potential $V^{(1)}$, the Hamiltonian matrix $\bm{H}^{(1)}$ is defined and new expansion coefficients $\bm{U}^{(1)}$ are built. The procedure is evolved self-consistently until the change in $\bm{P}^{(1)}$ falls below a certain threshold. The desired properties are subsequently calculated.} \label{fig:cpscf}
\end{wrapfigure}

Density-functional perturbation theory (DFPT) extends the framework of DFT to the calculation of response properties of the electronic density. DFPT can give access to a large variety of properties, depending on the type of external stimulus that generates the perturbation~\cite{baro+rmp2001}. An incomplete list of these properties includes force constant matrices and phonons~\cite{gian+prb1991}, molecular polarizabilities~\cite{maha+pra1980} and dielectric properties of solids~\cite{baro+rmp2001,gonz+prb1997}, vibrational scattering cross sections~\cite{raim+prm2019, putr+jcp2000}, electron-phonon coupling and non-adiabatic matrix elements~\cite{gius+prb2007, box+es2023}, NMR shifts and J-couplings~\cite{laasner+es2024}.

The implementation of these techniques within a numeric atom-centered orbital (NAO) framework and real-space context, aligned with the FHI-aims code architecture, necessitates specific choices that we outline in the next section. The original implementation, as detailed in papers by H. Shang and coworkers~\cite{shang+cpc2017,shang+njp2018}, employed independent routines for different perturbation scenarios. Recently, this approach has been streamlined by centralizing these routines, resulting in a more robust implementation and improved support for the code. \cite{box+ecse2021}

Like other functionalities of FHI-aims, both periodic and non-periodic evaluations are possible within the same all-electron, NAO infrastructure. For periodic phonons, only limited support (LDA functional, and systems that do not require large memory) is provided in the current implementation. A more comprehensive and very efficient handling of the calculation of phonons with different functionals and larger system sizes is provided through the FHI-aims connection to the FHI-vibes infrastructure~\cite{vibes}, detailed in another contribution of this Roadmap.

\section*{Current Status of the Implementation \\ \textit{\color{gray}(around 450 words)} }

In the following, we summarize the main aspects of the implementation regarding atomic displacement and electric field perturbations. A dedicated section on the implementation for magnetic field perturbations is given in another contribution of this Roadmap. 

In FHI-aims, the Kohn-Sham orbitals $\psi$ are expanded in local NAO basis
\begin{equation}
\psi^{(0)}_p = \sum_\mu C_{\mu p}^{(0)} \chi_\mu, \label{eq:psi0}
\end{equation}
where in the above notation $p$ is the orbital index, $\mu$ is an index running over the basis functions, the superscript $(0)$ indicates that this is a ground-state quantity, and $\chi$ are atom-centered basis functions in finite systems and Bloch-like superpositions of atom-centered basis functions in the periodic case. For the latter case, $\chi$ also carries a complex phase (see Eqs. 23 and 24 in Ref.~\cite{shang+njp2018}) and the orbitals would depend on $\bm{k}$, which we are not showing to simplify notation. Denoting first-order response quantities with the superscript $(1)$, standard first-order perturbation theory yields the Sternheimer equation~\cite{ster+pr1954} (for each $\bm{k}$ point)
\begin{equation}
(\hat{h}_{KS}^{(0)} - \epsilon_p^{(0)}) | \psi_p^{(1)} \rangle = -(\hat{h}^{(1)}_{KS} - \epsilon_p^{(1)}) | \psi_p^{(0)} \rangle.
\end{equation}
In order to solve it, we can expand the response of the wavefunction analogously to Eq.~\ref{eq:psi0},
\begin{equation}
\psi^{(1)}_p = \sum_\mu \left[ C^{(1)*}_{\mu p} \chi_\mu^{(0)}  + C^{(0)*}_{\mu p} \chi_\mu^{(1)}  \right] .\label{eq:psi1}
\end{equation}

For nuclear displacements, the atom-centered basis sets $\chi$  indeed change upon displacement and thus yield a non-zero $\chi_\mu^{(1)}$, whilst in the case of an electric-field perturbation this term is zero and Eq.~\ref{eq:psi1} simplifies to only the first term. While DFPT formalisms concentrate on a self-consistent procedure to determine directly the coefficients $C^{(1)}$, the procedure commonly named coupled perturbed self-consistent field (CPSCF) writes $C^{(1)}_{\mu p} = \sum_q U_{pq} C_{\mu q}^{(0)} $ as a linear expansion on the unperturbed coefficients. The new coefficients $U_{pq}$, where $p$ denotes an occupied orbital and $q$ an unoccupied one, take the following form
\begin{equation} \label{eq:u1}
 U_{pq} =  \begin{cases}
     \frac{\sum_{\mu \nu}(C_{\mu q}^{(0)})^{*}H_{\mu\nu}^{(1)} C_{\nu p}^{(0)}}{\epsilon_p^{(0)} - \epsilon_tq^{(0)}}, & \text{el. field}\\
    \frac{\sum_{\mu \nu}(C_{\mu q}^{(0)})^{*}H_{\mu\nu}^{(1)} C_{\nu p}^{(0)} - \sum_{\mu \nu} (C_{\mu q}^{(0)})^{*} S_{\mu\nu}^{(1)} (\bm{C}^{(0)} \bm{E}^{(0)})_{\nu p}}{\epsilon_p^{(0)} - \epsilon_q^{(0)}}, & \text{nuclear displ.}
  \end{cases}
\end{equation}
and a self-consistent procedure is also necessary to determine them (see Fig.~\ref{fig:cpscf}). Note that the response of the overlap matrix ($\bm{S}^{(1)}$) is only necessary for nuclear displacement perturbations. The CPSCF formulation, while completely equivalent to DFPT, turns out to be advantageous in FHI-aims because it allows us to make use of existing algorithms for the massively parallel evaluation of matrix elements in this representation. This expansion also allows computing the density response
\begin{eqnarray}
n^{(1)} & = &  \sum_p f(\epsilon_p) \left[\psi^{(1)*}\psi^{(0)} + \psi^{(0)*}\psi^{(1)}\right] = \\ 
 & = & \begin{cases} 
 \sum_{\mu \nu} P_{\mu \nu}^{(1)} \chi_{\mu}^{(0)} \chi_{\nu}^{(0)},  &  \text{el. field} \\
 \sum_{\mu \nu} [P_{\mu \nu}^{(1)} \chi_{\mu}^{(0)} \chi_{\nu}^{(0)} + P_{\mu \nu}^{(0)}(\chi_{\mu}^{(1)} \chi_{\nu}^{(0)} + \chi_{\mu}^{(0)} \chi_{\nu}^{(1)}) ] & \text{nuclear displ.}
  \end{cases}
\end{eqnarray}
within a response density-matrix $\bm{P}^{(1)}$ formalism, analogous to the ground-state density evaluation in the code. The full self-consistent procedure is summarized in Fig.~\ref{fig:cpscf}.

\begin{figure}[ht]
    \centering
    \includegraphics[width=\textwidth]{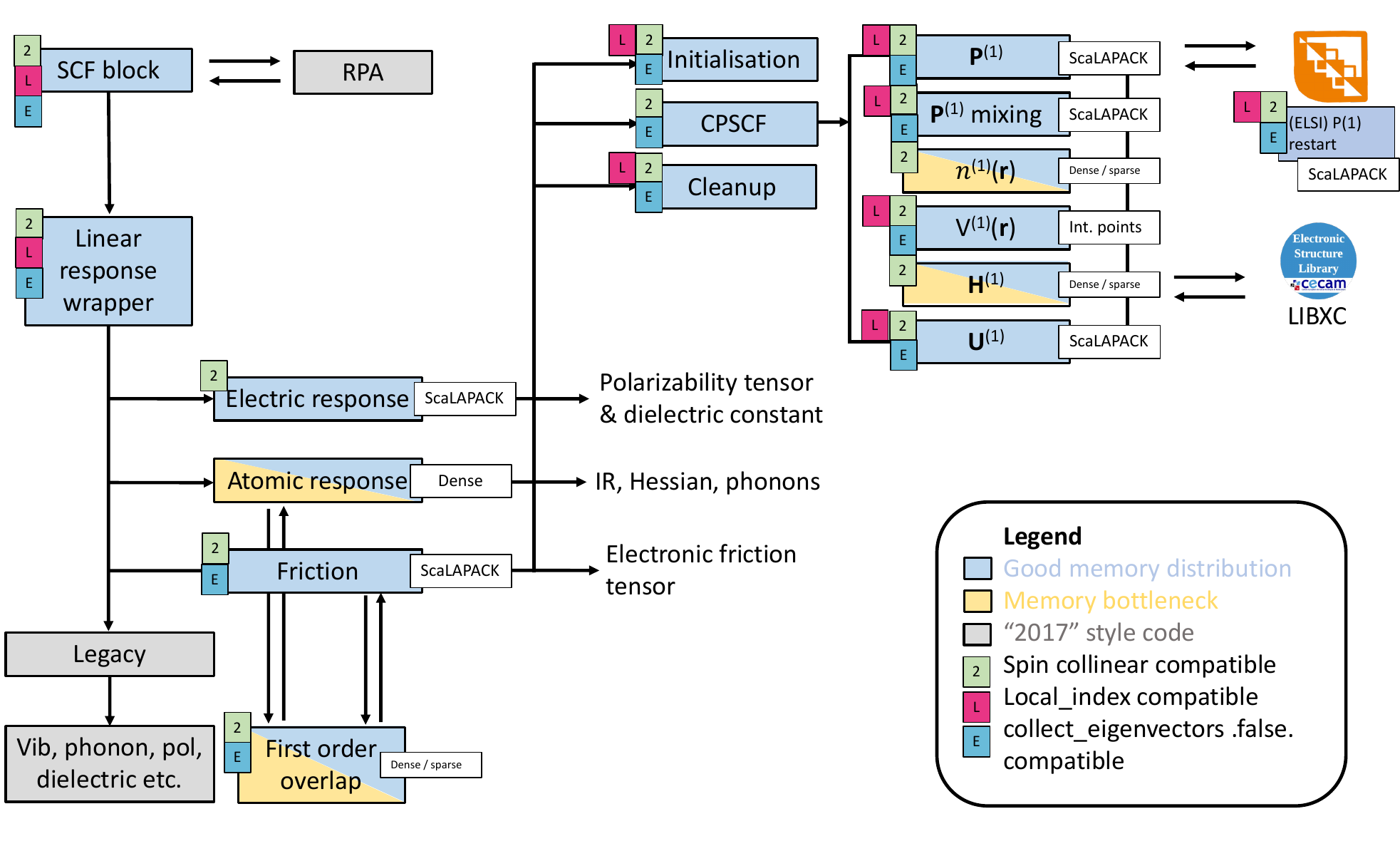}
    \caption{Schematic representation of the current code structure in FHI-aims. ``Dense / sparse'' refers to when the full global dense (aperiodic) or sparse (periodic) response matrices are needed, ``Int. points'' refers to distribution over integration grid points and ``ScaLAPACK'' refers to ScaLAPACK support for distribution of matrices. Mixed coloured regions currently represent a memory bottleneck for aperiodic systems due to the use of full dense matrices.}
    \label{fig:fhiaims-flow}
\end{figure}

An overview of the current implementation is given in Fig.~\ref{fig:fhiaims-flow}. In this implementation, the DFPT calculations are governed by a central ``linear response wrapper'' function that is called after convergence of the ground-state DFT calculation. This function can call different application modules, which currently include ``electric response'' (for polarisability, dielectric constant) ``atomic response'' (for phonon spectra and molecular vibrations), and ``electronic friction'' (for electron-phonon coupling) calculations. This portfolio of options can easily be further extended, as this infrastructure provides a well documented template for future application modules employing the central DFPT infrastructure. An example for a future extension of the central DFPT framework is given in Fig.~\ref{fig:fhiaims-flow} as the random phase approximation (RPA) functionality, which requires certain response quantities to compute forces. Recently, we have extended the support to allow  ``electric response'' and ``electronic friction'' calculations of fractionally occupied systems (including metallic systems) which necessities special treatment of the denominator in Eq.~\ref{eq:u1}.

The central CPSCF module now features interfaces to external open-source libraries such as ELSI~\cite{yu+cpc2020} and LibXC~\cite{marques+cpc2012}. ELSI enables a restart of the CPSCF routine by parallelised I/O of the first order density matrix and LibXC enables the calculation of the first order Hamiltonian for arbitrary LDA, GGA, and (when supported) hybrid-exchange-correlation functionals. 

Routines that represent significant bottlenecks in memory distribution and parallelisation are shaded yellow in Fig.~\ref{fig:fhiaims-flow}, whilst routines with optimised memory distribution are shaded blue. The half-shading refers to having full dense matrices allocated in the aperiodic case which is memory inefficient when compared to cases that support ScaLAPACK-type distribution, whilst real-space sparse matrices are allocated in the periodic case, which is generally a memory efficient approach. For the latter case, \texttt{use\_local\_index} may be used which reduces memory usage and can improve efficiency, particularly for large systems, routines that support this are labelled.  The figure also depicts routines that currently support 2 spin channels (i.e spin collinear) and \texttt{collect\_eigenvectors .false.} (for when ScaLAPACK-type handling of matrix operations is employed, this significantly reduces memory usage).

\section*{Usability and Tutorials \\ \textit{\color{gray}(around 400 words)}}

All functionality is now accessible through simple keywords in the \texttt{control.in} file, documented in the FHI-aims manual since release 240507. 
Currently, there are keywords that control the central CPSCF shared by all driver routines, these all start with the prefix \texttt{dfpt\_} and control accuracy thresholds, mixing and restart behaviour.
The actual DFPT calculation is triggered by one of the driver related keywords, currently these are \texttt{electric\_field\_response~DFPT}, \texttt{atomic\_pert\_response~DFPT} or \texttt{calculate\_friction~DFPT}.

\begin{figure}[ht]
    \centering
    \includegraphics[width=\textwidth]{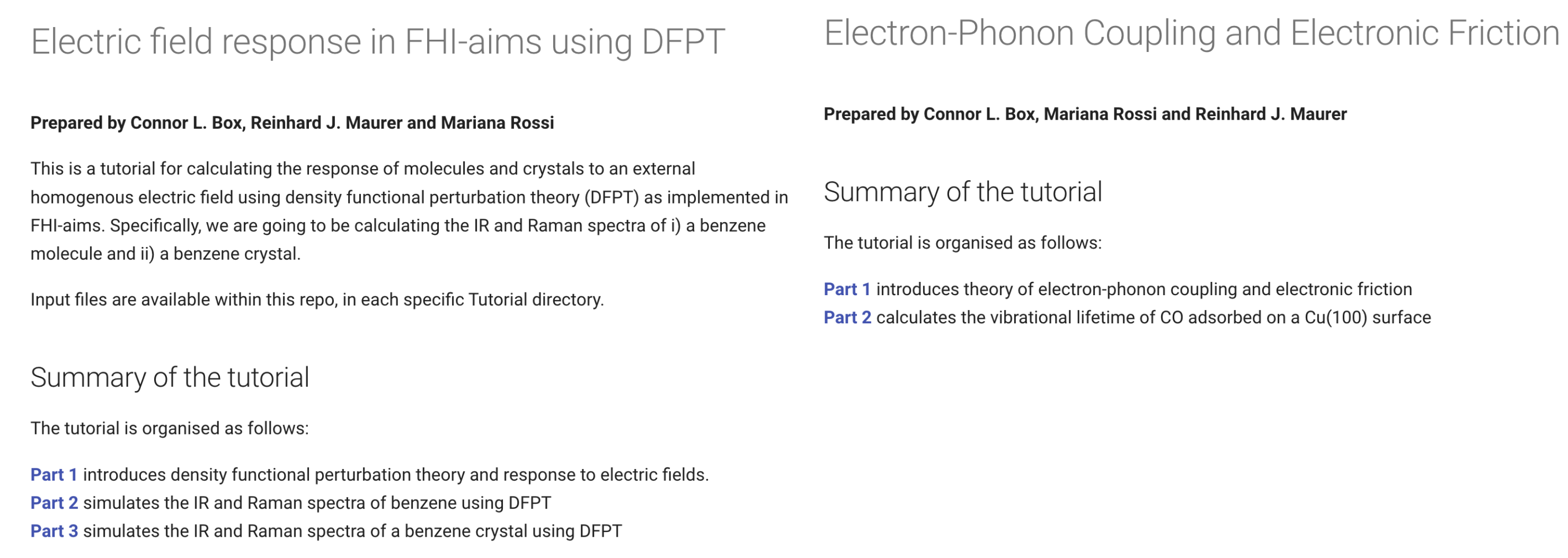}
    \caption{Screenshot of tutorials showcasing the usability and utility of DFPT in FHI-aims.}
    \label{fig:tutorials}
\end{figure}

We have recently released online tutorials, which will be continuously updated and augmented, in order to make the functionality more accessible to new users and showcase what the DFPT functionality can achieve. One tutorial employs the electric-field response functionality to calculate non-resonant vibrational Raman spectra of an isolated molecule and a molecular crystal. Notably, this tutorial also showcases the new implementation of the Berry-phase polarization through the use of Wannier orbitals (showcased in another article of this Roadmap) in order to calculate IR spectra of periodic systems. Another tutorial makes use of the electronic friction driver to calculate electron-phonon couplings and vibrational lifetimes including non-adiabatic effects through electronic friction for a molecular overlayer adsorbed on a metallic surface. These effects are discussed in depth in another contribution of this Roadmap. The landing page and brief description of both tutorials are showcased in Fig.~\ref{fig:tutorials}. These tutorials can be accessed through \url{https://fhi-aims-club.gitlab.io/tutorials/tutorials-overview/}.

\section*{Future Plans and Challenges \\ \textit{\color{gray}(around 350 words)}}

The newly refactored infrastructure~\cite{box+ecse2021} is significantly more compact and easier to maintain than the original. The total number of code lines has been reduced by just over 60\% for the CPSCF code, with the newer interface also including more features. Several existing versions of multiple routines have been consolidated, reducing code duplication and increasing clarity for future developers.
Specific run-mode cases are now only dealt with at the lowest level and high level developments at the level of the CPSCF cycle are decoupled. In the near future, we plan to extend the support of local dense matrix computations and the distribution of eigenvectors (currently, a copy of the full eigenvector is created on every MPI task for the $\bm{k}$-point(s) it is working on). These changes will improve memory usage and computational efficiency of not just the CPSCF calculations but allow these features to be utilised in the SCF part of the calculation as well, where they are already supported. A scalable implementation of DFPT based on all-electron NAO basis that performs across various HPC systems has been reported~\cite{Shang2021}. These implementations can address systems up to 200,000 atoms. These developments are compatible with FHI-aims.

It is also important to stress the role that machine-learning (ML) plays in the calculation of electronic response properties. While this field is still less advanced than other topics of ML for material-science, FHI-aims is already able to produce the density-response data that is needed to train models that learn the electronic density response directly, or its derived quantities. The success of these frameworks based on FHI-aims data in predicting Raman spectra, dielectric properties of materials and electronic friction tensors has already been established~\cite{raim+njp2019,zhan+jpcc2020,shan+aipa2021, lewi+jcp2023}. This machinery can massively speed up computations and is expected to have an even larger impact on the calculation of electronic response properties in the coming years.

\section*{Acknowledgements}
We acknowledge fruitful discussions with Andrew Logsdail, Ville Havu, Xinguo Ren (and others). We acknowledge financial support through the UKRI Future Leaders Fellowship programme (MR/S016023/1 and MR/X023109/1), and a UKRI frontier research grant (EP/X014088/1).

\end{document}